\documentstyle[sprocl,epsfig]{article}

\def\Journal#1#2#3#4{{#1} {\bf #2}, #3 (#4)}

\def\NCA{{\em Nuovo Cimento} A}

\def\NIMA{{\em Nucl. Instrum. Methods} A}
\def\NPB{{\em Nucl. Phys.} B}
\def\PLB{{\em Phys. Lett.}  B}

\def\ZPC{{\em Z. Phys.} C}

\def\be{\begin{equation}}
\def\ee{\end{equation}}
\def\bea{\begin{eqnarray}}
\def\eea{\end{eqnarray}}
\newcommand{\pipipipi}{\mbox{$\pi^+\pi^-\pi^+\pi^-$ }}

\newcommand{\pipi}{\mbox{$\pi^{+}\pi^{-}$} }

\newcommand{\fmeson}{\mbox{$f_{2}(1270)$} }

\begin{document}
\title{NEW EFFECTS OBSERVED IN CENTRAL PRODUCTION BY THE WA102 EXPERIMENT AT
THE CERN OMEGA SPECTROMETER}

\author{Andrew Kirk$^*$ \\
representing the WA102 collaboration$^{1}$}
\address{$^*$School of Physics and Astronomy, Birmingham University, U.K.\\
E-mail: andrew.kirk@cern.ch}

\maketitle

\abstracts{
A study of central meson production as a function of the
difference in transverse momentum ($dP_T$)
of the exchanged particles
shows that undisputed $q \overline q$ mesons
are suppressed at small $dP_T$ whereas the glueball candidates
are enhanced.
In addition, the production cross section for different
resonances depends strongly on the azimuthal angle between the
two outgoing protons.
}

\section{Introduction}
\par
There is considerable current interest in trying to isolate the lightest
glueball.
Several experiments have been performed using glue-rich
production mechanisms.
One such mechanism is Double Pomeron Exchange (DPE) where the Pomeron
is thought to be a multi-gluonic object.
Consequently it has been
anticipated that production of
glueballs may be especially favoured in this process~\cite{closerev}.
\par
The Omega central production experiments
(WA76, WA91 and WA102) are
designed to study exclusive final states
formed in the reaction
\noindent
\be
pp \longrightarrow p_{f} X^{0} p_s,
\label{eq:1}
\ee
where the subscripts $f$ and $s$ refer to the fastest and slowest
particles in the laboratory frame respectively and $X^0$ represents
the central system. Such reactions are expected to
be mediated by double exchange processes
where both Pomeron and Reggeon exchange can occur.
\par
The trigger was designed to enhance double exchange
processes with respect to single exchange and elastic processes.
Details of the trigger conditions, the data
processing and event selection
have been given in previous publications~\cite{re:expt}.
\section{A Glueball-$q \overline q$ filter in central production ?}
\par
The experiments have been
performed at incident beam momenta of 85, 300 and 450 GeV/c, corresponding to
centre-of-mass energies of
$\sqrt{s} = 12.7$, 23.8 and 28~GeV.
Theoretical
predictions \cite{pred} of the evolution of
the different exchange mechanisms with centre
of mass energy, $\sqrt{s}$, suggest that
\noindent
\bea
 \sigma (RR)  \sim s^{-1} , \nonumber \\
 \sigma (RP)  \sim s^{-0.5} , \nonumber \\
 \sigma (PP)  \sim  constant,
\label{eq:2}
\eea
where RR, RP and PP refer to Reggeon-Reggeon, Reggeon-Pomeron and
Pomeron-Pomeron
exchange respectively. Hence we expect Double Pomeron Exchange
(DPE) to be more significant at high energies, whereas the Reggeon-Reggeon and
Reggeon-Pomeron mechanisms will be of decreasing importance.
The decrease of the non-DPE cross section with energy can be inferred
from data
taken by the WA76 collaboration using pp interactions at $\sqrt{s}$ of 12.7 GeV
and 23.8 GeV \cite{wa76}.
The \pipi mass spectra for the two cases show that
the signal-to-background ratio for the $\rho^0$(770)
is much lower at high energy, and the WA76 collaboration report
that the ratio of the $\rho^0$(770) cross sections at 23.8 GeV and 12.7 GeV
is 0.44~$\pm$~0.07.
Since isospin 1 states such as the $\rho^0$(770) cannot be produced by DPE,
the decrease
of the $\rho^{0}(770)$ signal at high $\sqrt{s}$
is consistent with DPE becoming
relatively more important with increasing energy with respect to other
exchange processes.
\par
\begin{figure}[h]
\begin{center}
\epsfig{file=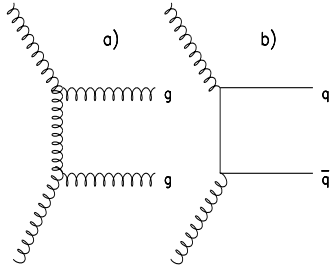,height=5.0cm,width=4.5cm,
bbllx=0pt,bblly=0pt,bburx=550pt,bbury=500pt}
\end{center}
\caption{Schematic diagrams
of the coupling of the exchange particles into the final state meson
for a) gluon exchange and b) quark exchange.}
\label{fi:feyn1}
\end{figure}

However,
even in the case of pure DPE
the exchanged particles still have to couple to a final state meson.
The coupling of the two exchanged particles can either be by gluon exchange
or quark exchange. Assuming the Pomeron
is a colour singlet gluonic system if
a gluon is exchanged then a gluonic state is produced, whereas if a
quark is exchanged then a $q \overline q $ state is produced
(see figures~\ref{fi:feyn1}a) and b) respectively).
\begin{figure}[b!]
\begin{center}
\epsfig{file=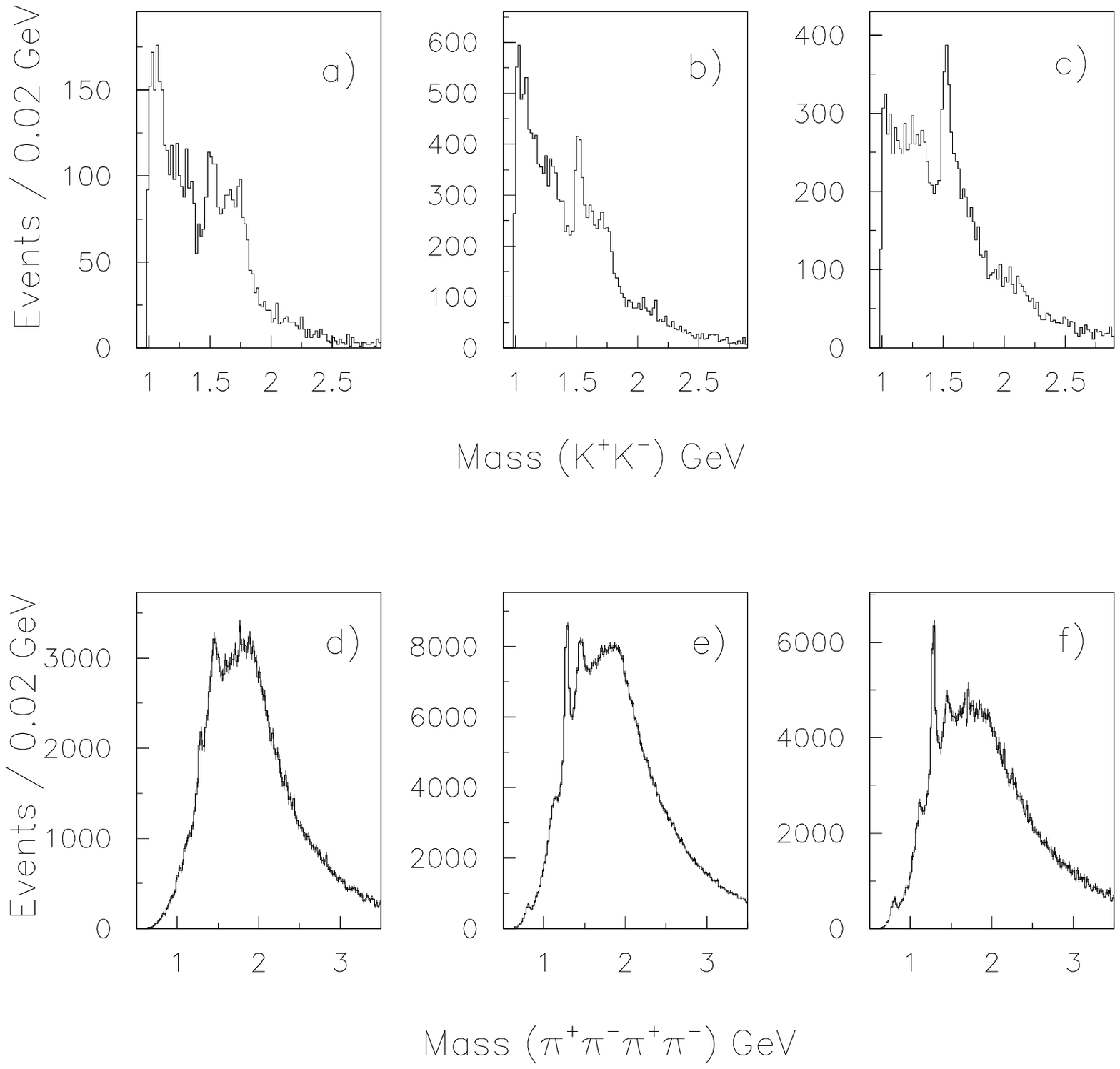,height=12cm,width=14cm}
\end{center}
\caption{$K^+K^-$ mass spectrum for a) $dP_T <   0.2$ GeV,
b) $0.2 <   dP_T <   0.5$ GeV and c) $dP_T >   0.5$ GeV and
the \pipipipi mass spectrum for d) $dP_T <   0.2$ GeV,
e) $0.2 <   dP_T <   0.5$ GeV and f) $dP_T >   0.5$ GeV.}
\label{fi:2k}
\end{figure}
\par
The WA91 collaboration has published a paper~\cite{wa91corr}
showing that the observed centrally produced resonances depend on the
angle between the outgoing slow and fast protons.
In order to describe the data in terms of a physical model,
Close and Kirk~\cite{closeak},
have proposed that the data be analysed
in terms of the difference in transverse momentum
between the particles exchanged from the
fast and slow vertices.
The idea being that
for small differences in transverse momentum between the two
exchanged particles
an enhancement in the production of glueballs
relative to $q \overline q$ states may occur.
The difference in the transverse momentum vectors ($dP_T$) is defined to be
\be
dP_T = \sqrt{(P_{y1} - P_{y2})^2 + (P_{z1} - P_{z2})^2}
\label{eq:3}
\ee
where
$Py_i$, $Pz_i$ are the y and z components of the momentum
of the $ith$ exchanged particle in the pp centre of mass system~\cite{dptpap}.
\begin{figure}
\begin{center}
\epsfig{file=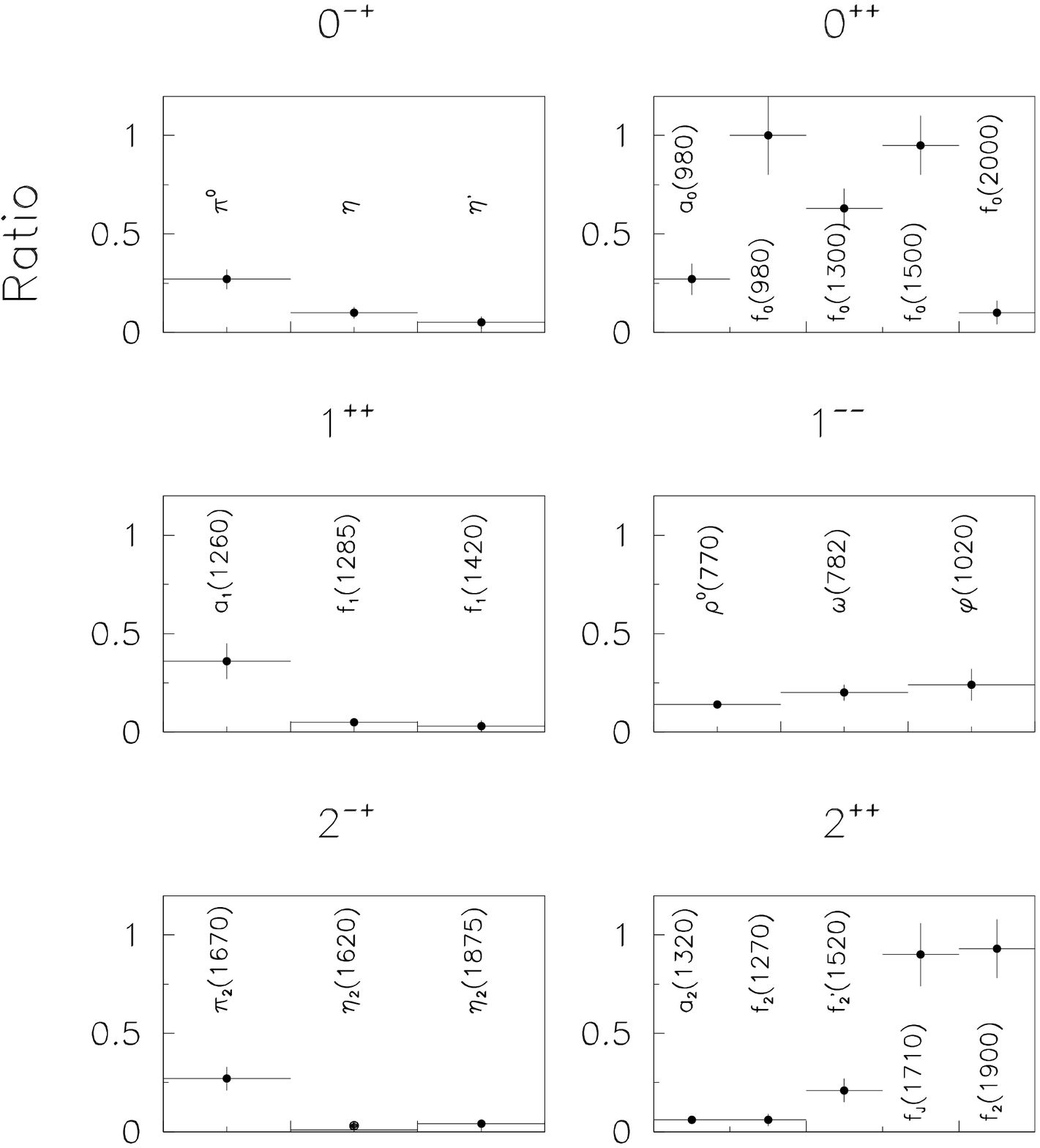,height=16cm,width=12cm}
\end{center}
\caption{The ratio of the amount of resonance with
$dP_T$~$\leq$~0.2 to the amount with
$dP_T$~$\geq$~0.5~GeV.
}
\label{fracratio}
\end{figure}
\par
Figures~\ref{fi:2k}a), b) and c) show the effect of the
$dP_T$ cut on the $K^+ K^-$ mass spectrum where
structures can be observed in the 1.5 and 1.7 GeV mass region which have
been previously identified as the
$f_{2}^\prime$(1525) and the $f_J(1710)$~\cite{re:WA76KK}.
As can be seen,
the $f_{2}^\prime$(1525) is produced dominantly at high $dP_T$,
whereas the $f_J(1710)$ is produced dominantly at low $dP_T$.
\par
In the \pipipipi  mass spectrum a dramatic effect is observed,
see figures~\ref{fi:2k}d), e) and f).
The $f_1(1285)$ signal has virtually disappeared at low $dP_T$
whereas
the $f_0(1500)$ and $f_2(1930)$ signals remain.
\par
In fact it has been observed~\cite{memoriam} that
all the undisputed
$ q \overline q $ states
(i.e. $\rho^0(770)$, $\eta^{\prime}$, \fmeson, $f_1(1285)$,
$f_2^\prime(1525)$ etc.)
are suppressed as $dP_T$ goes to zero,
whereas the glueball candidates
$f_J(1710)$, $f_0(1500)$ and $f_2(1930)$ survive.
It is also interesting to note that the
enigmatic
$f_0(980)$,
a possible non-$q \overline q$ meson or $K \overline K$ molecule state does not
behave as a normal $q \overline q$ state.
\par
A Monte Carlo simulation of the trigger, detector acceptances
and reconstruction program
shows that there is very little difference in the acceptance as a function of
$dP_T$ in the different mass intervals considered
within a given channel and hence the
observed differences in resonance production can not be explained
as acceptance effects.
\section{Summary of the effects of the $dP_T$ filter}
\par
In order to calculate the contribution of each resonance as a function
of the $dP_T$ the mass spectra have been fitted with
the parameters of the resonances fixed to those obtained from the
fits to the total data.
Figure~\ref{fracratio} shows the ratio of the number of events
for $dP_T$ $<$ 0.2 GeV to
the number of events
for $dP_T$ $>$ 0.5 GeV for each resonance considered.
It can be observed that all the undisputed $q \overline q$ states
which can be produced in DPE, namely those with positive G parity and $I=0$,
have a very small value for this ratio ($\leq 0.1$).
Some of the states with $I=1$ or G parity negative,
which can not be produced by DPE,
have a slightly higher value ($\approx 0.25$).
However, all of these states are suppressed relative to the
interesting states, i.e. those which could have a gluonic component, which have
a large value for this ratio.
\begin{figure}
\begin{center}
\epsfig{file=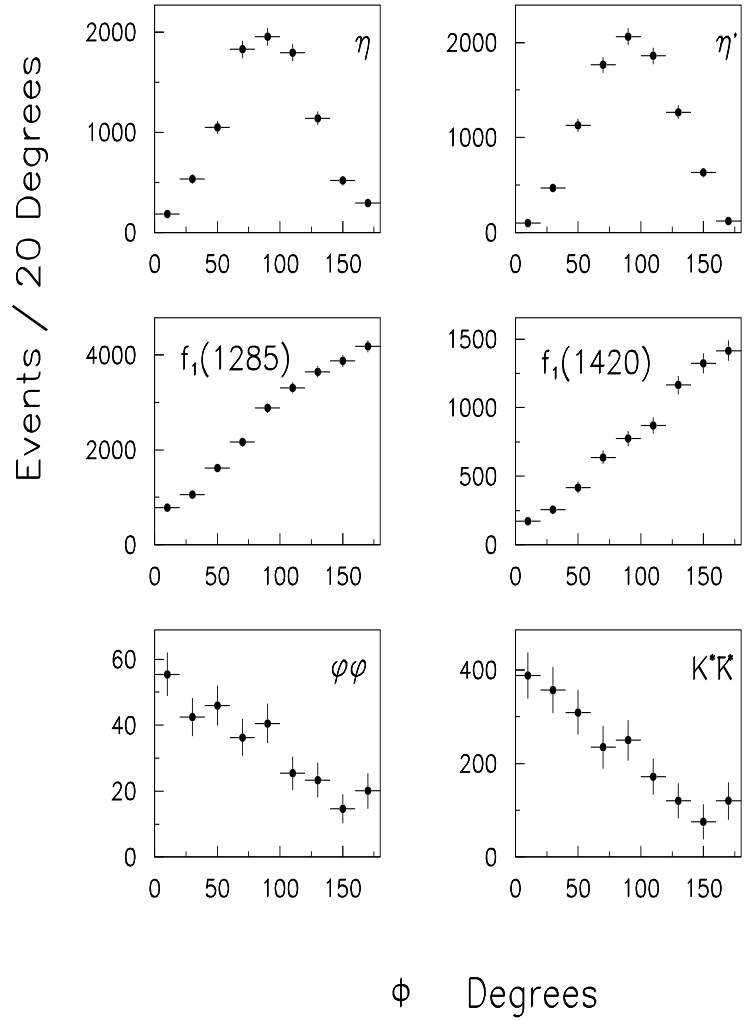,height=16cm,width=12cm}
\end{center}
\caption{The azimuthal angle between the fast and
slow protons ($\phi$) for various final states.
}
\label{fi:phidep}
\end{figure}

\section{The azimuthal angle between the outgoing protons}

\par
The azimuthal angle ($\phi$) is defined as the angle between the $p_T$
vectors of the two protons.
Naively it may be expected that this angle would be flat irrespective
of the resonances produced.
Fig.~\ref{fi:phidep}  shows the $\phi$ dependence for two
$J^{PC}$~=~$0^{-+}$ final states (the $\eta$ and $\eta^\prime$),
two $J^{PC}$~=~$1^{++}$ final states (the $f_1(1285)$ and $f_1(1420)$) and
two $J^{PC}$~=~$2^{++}$ final states
(the $\phi \phi$ and $K^*(892) \overline K^*(892)$ systems).
The $\phi$ dependence is clearly not flat and considerable variation
is observed between final states with different $J^{PC}$s.
\section{Implications of the $dP_T$ and azimuthal angle effects}
\par
The underlying physics behind the $dP_T$ and azimuthal angle ($\phi$) effects
is still not fully understood. They are not an artifact of the WA102 experiment
since they have been subsequently verified by the NA12/2 experiment.
The angle $\phi$ is related to $dP_T$ by
\be
cos \phi  = \frac{ dP_T^2  -  P_T^2}{4 t_s t_f}
\label{eq:4}
\ee
where $P_T$ is the transverse momentum of the central system.
It is not possible, however, based on this relation alone to
explain the  $\phi$ distribution for the $\eta$ and $\eta^prime$.
\par
It has been suggested that it may be possible to explain the results if the
particles exchanged in the formation of the central system carry non-zero
spin~\cite{closedpe,frere}. Hence the results may have implications
for the spin structure of the Pomeron.
\section{Conclusions}
\par
A study of centrally produced pp interactions
show that there is the possibility of a
glueball-$q \overline q$ filter mechanism.
All the
undisputed $q \overline q $ states are observed to be suppressed
at small $dP_T$, but the glueball candidates
$f_0(1500)$, $f_J(1710)$, and $f_2(1930)$ ,
together with the enigmatic $f_0(980)$,
survive.
In addition, the production cross section for different
resonances depends strongly on the azimuthal angle between the
two outgoing protons.

\end{document}